\documentclass[fleqn,10pt]{wlscirep}
\usepackage[utf8]{inputenc}
\usepackage[T1]{fontenc}
\title{Out-of-equilibrium charge redistribution in a copper-oxide based superconductor by time-resolved X-ray photoelectron spectroscopy}

\author[1] {Denny Puntel}
\author[2] {Dmytro Kutnyakhov}
\author[2] {Lukas Wenthaus} 
\author[2] {Markus Scholz} 
\author[2,3] {Nils O. Wind} 
\author[2] {Michael Heber} 
\author[2] {G\"{u}nter Brenner} 
\author[4]{Genda Gu}
\author[5]{Robert J. Cava}
\author[6] {Wibke Bronsch}
\author[6] {Federico Cilento}
\author[1,6,7*] {Fulvio Parmigiani}
\author[2,**] {Federico Pressacco}

\affil[1]{Dipartimento di Fisica, Università degli Studi di Trieste, Trieste, 34127, Italy}
\affil[2]{Deutsches Elektronen-Synchrotron DESY, Hamburg, 22607, Germany}
\affil[3]{University of Hamburg, Institut für Experimentalphysik, Hamburg, 22761, Germany}
\affil[4]{Condensed Matter Physics and Materials Science Department, Brookhaven National Laboratory, Upton, New York 11973, USA}
\affil[5]{Department of Chemistry, Princeton University, Princeton, New Jersey 08544, USA}
\affil[6]{Elettra - Sincrotrone Trieste S.C.p.A., Trieste, 34149, Italy}
\affil[7]{International Faculty, University of Cologne, Cologne, 50923, Germany}

\affil[*]{fulvio.parmigiani@elettra.eu}
\affil[**]{federico.pressacco@desy.de}

\begin{abstract}
Charge-transfer excitations are of paramount importance for understanding the electronic structure of copper-oxide based high-temperature superconductors. In this study, we investigate the response of a Bi$_2$Sr$_2$CaCu$_2$O$_{\mathrm{8}+ \delta}$ crystal to the charge redistribution induced by an infrared ultrashort pulse. Element-selective time-resolved core-level photoelectron spectroscopy with a high energy resolution allows disentangling the dynamics of oxygen ions with different coordination and bonds thanks to their different chemical shifts. Our experiment shows that the O\,$1s$ component arising from the Cu-O planes is significantly perturbed by the infrared light pulse. Conversely, the apical oxygen, also coordinated with Sr ions in the Sr-O planes, remains unaffected. This result highlights the peculiar behavior of the electronic structure of the Cu-O planes. It also unlocks the way to study the out-of-equilibrium electronic structure of copper-oxide-based high-temperature superconductors by identifying the O\,$1s$ core-level emission originating from the oxygen ions in the Cu-O planes. This ability could be critical to gain information about the strongly-correlated electron ultrafast dynamical mechanisms in the Cu-O plane in the normal and superconducting phases.
\end{abstract}
\begin{document}

\flushbottom
\maketitle
%
%
\thispagestyle{empty}

\section*{Introduction}

After more than thirty years of study, the mechanism that induces a superconducting state in some layered copper oxides remains obscure. Nonetheless, much information has accumulated. In particular, we know that many-body interactions in the Cu-O planes are the key to elucidating the process of Cooper-pair formation and condensation into a superconductive state at temperatures that cannot be accounted for by a model that attributes the superconductivity to the electron-phonon coupling \cite{Giannetti2011, Damascelli2003, Basov2005, Lee2006}. A spectroscopy capable of revealing the ultrafast dynamics of the strongly correlated electrons into the Cu-O plane could provide vital information to solve the problem of why superconductivity can occur at such high temperatures. It has been shown that X-ray photoelectron spectroscopy (XPS) of the O\,$1s$ core-level structure can resolve emissions that depend on the oxygen bonds and coordination  \cite{Hill1988, Hinnen1995, Kohiki1988, Soederholm1996, Kuo2018}, however technical aspects such as limited signal statistics and energy resolution have so far hindered its further applications to cuprates.

Here we show that with an element- and coordination-sensitive probe it is possible to disentangle the dynamics of the charge redistribution processes induced by a $\sim 70$\,fs infrared light pulse in a cuprate superconductor. In particular, we report on the first sub-picosecond time-resolved XPS study on optimally-doped Bi$_2$Sr$_2$CaCu$_2$O$_{8+\delta}$ (Bi2212) above the superconducting transition temperature, to unveil the response of the oxygen and strontium ions far from the equilibrium conditions. We prove that the effects induced by a charge redistribution in the valence band of Bi2212 can be observed by inspection of the core-level dynamics. Thanks to a time-resolved XPS experiment with improved energy resolution, our measurements reveal marked differences in the ultrafast dynamics of the three components contributing to the O1s line emission. In particular, the oxygen in the Cu-O plane is the most affected by photoexcitation, as signaled by a shift and a broadening of the corresponding spectral component. The effects experienced by the oxygen in the Sr-O plane are instead negligible, as also confirmed by inspection of the Sr\,$3d$ emission. This result indicates that the response of cuprate systems to the charge redistribution induced by the infrared pump pulse predominantly involves the Cu-O plane. \\

\begin{figure}[ht]
\includegraphics[width=\columnwidth]{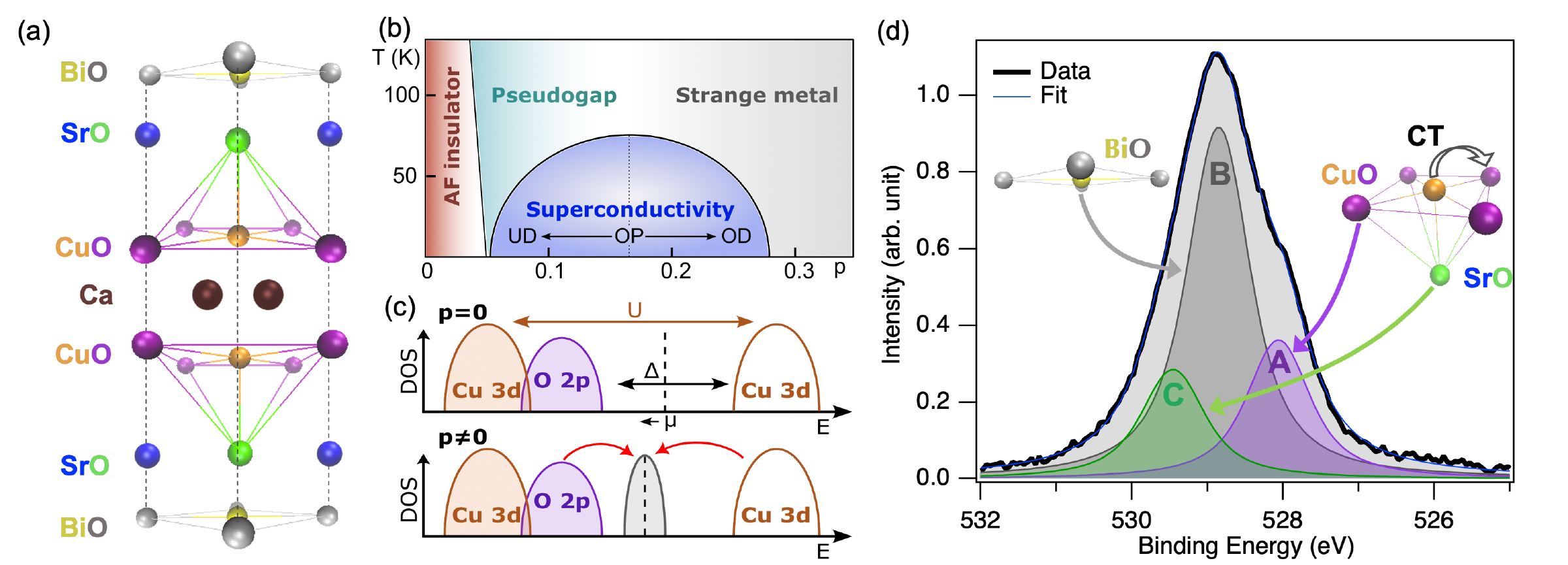}
\caption{(\textbf{a}) Crystal structure of $\mathrm{Bi_2Sr_2CaCu_2O_8} $ (Bi2212). Oxygen atoms occupying different planes are indicated with different colors. (\textbf{b}) Phase diagram of Bi2212 as a function of hole doping (p, hole per Cu site) and temperature, in the low-temperature region. The position of optimal doping, where the highest $\mathrm{T_C}$ is reached, is marked by a dotted line ($\mathrm{p} = 0.16$). (\textbf{c}) Schematic of the density of states close to the chemical potential in an undoped charge-transfer insulator ($\mathrm{p} = 0$) and in the doped conducting system ($\mathrm{p} \neq 0$) reminiscent of the insulating structure. U, Mott-Hubbard gap between Cu states; $\Delta$, charge-transfer gap; $\mu$, chemical potential. (\textbf{d}) O\,$1s$ emission from an optimally-doped Bi2212 acquired at FLASH at 592\,eV photon energy and temperature of 100\,K. The fit (see main text) is superimposed as a blue line. The spectrum shows the envelope of three components experiencing different chemical shifts induced by the environment, which allows assigning each component to oxygen atoms in a specific layer of the unit cell: component A (violet) to planar Cu-O oxygens, component B (grey) to Bi-O oxygens, component C (green) to apical Sr-O oxygens. 
\label{fig1}}
\end{figure}

A schematic crystal structure of $\mathrm{Bi_2Sr_2CaCu_2O_8}$ (Bi2212) is shown in Fig.~\ref{fig1}a. It comprises several oxide planes stacked along the $c$ axis, with two copper-oxygen planes per unit cell. The unit cells are held together by van-der-Waals forces among the Bi-O planes,  which makes it possible to exfoliate the Bi2212 samples along the $c$ axis, with Bi-O as an exposed plane \cite{Kirk1988, Lindberg1989}. Although the superconducting character of cuprates is mostly determined by the many-body interactions within the Cu-O planes, the presence of oxygen in the other planes also plays a significant role. The phonons associated with the motion of the oxygen in the Sr-O plane (apical oxygen) have a dramatic impact on superconductivity \cite{Pavarini2001, Hu2014, Mankowsky2014, Liu2020}. Moreover, additional holes are introduced in the Cu-O plane upon variation of the oxygen content in the remaining layers (thus called the \emph{charge reservoir} layers) defining the landscape of phases illustrated in the low-temperature phase diagram of Fig.~\ref{fig1}b. One of the most striking aspects of cuprate physics is that the parent undoped compound is an antiferromagnetic insulator (red area) and becomes conducting upon hole doping \cite{Lee2006}. The so-called superconducting dome (blue area) has a maximum critical temperature at the optimal doping value of $\mathrm{p} \simeq 0.16$, where $\mathrm{p}$ is the number of holes per Cu site. Upon increasing the temperature at moderate (under-doped, UD) or optimal (OP) doping values, the superconducting phase evolves into a second gapped phase, called pseudogap. The overdoped side of the phase diagram (OD) is occupied by a strange metal phase as the Fermi-liquid model fails to correctly predict its properties \cite{Varma1997, Fournier2010}.

The evolution from the insulating to the conducting state is schematized in Fig.~\ref{fig1}c. The large Coulomb repulsion experienced by the Cu\,$3d$ orbitals opens a correlation gap (U) which is larger than the bandwidth and would define the system as a Mott-Hubbard insulator. However, given the large U, the states at lowest energy are the O\,\emph{2p} ones, hence the first excitation is a charge transfer across the gap of amplitude $\Delta$ among oxygen and copper states in the valence and conduction bands respectively \cite{Ghijsen1990, Qvarford1996}. For this reason, the undoped cuprate compounds are more appropriately defined as a charge-transfer insulator \cite{Zaanen1985}. Upon doping, new states appear inside the gap at the down-shifted chemical potential (lower panel in Fig.~\ref{fig1}c) \cite{Romberg1990, Ohta1992}, but the remaining structure is still reminiscent of the charge-transfer phase \cite{Keimer2015, Peli2017, Kohsaka2007}. This physics leads to the intertwining of high-($1-10$\,eV) and low-energy scales ($<1$\,eV) typical of strongly-correlated systems \cite{Kuiper1989, Eskes1991, Meinders1993, Cooper1990}. Several pieces of evidence point out this interplay for cuprates \cite{Chen1991, Hybertsen1992, Basov2005, Giannetti2011, Baykusheva2022}. This indicates that also in the finite-doping phases, electron repulsion and charge transfer are the dominant interactions, thus influencing the low-energy physics of superconductivity and pseudogap (both gaps are of the order of tens of meV). However, the mechanism for which these high-energy interactions are at the origin of low-energy electrodynamics mechanisms still needs to be elucidated. The topic was investigated by time- and angle-resolved photoelectron spectroscopy by Cilento \emph{et al.} \cite{Cilento2018}. Led by the idea that a laser pulse of suitable photon energy can trigger the optical transition across the remnant charge-transfer structure, a 1.6\,eV pulse was used to redistribute the population from the occupied O\,\emph{2p} to the empty Cu\,$3d$ bands in the superconducting phase of an optimally-doped Bi2212 crystal, detecting the subsequent response of the band structure. New electronic states were observed to appear inside the gap as an effect of photoexcitation, along with a broadening of the O\,\emph{2p} band. It was found that the relaxation time of these two effects is identical, hence pointing to a direct link between the charge-transfer excitations from oxygen to copper states and the onset mechanism of the superconducting phase. 

A recent study on La$_{1.905}$Ba$_{0.095}$CuO$_4$ employed the chemical sensitivity provided by time-resolved X-ray absorption spectroscopy to detect a photoinduced renormalization of the effective Coulomb repulsion giving rise to the Hubbard gap \cite{Baykusheva2022}, hence renewing the interest in the role of high-energy excitations in determining the response of cuprate systems to photoexcitation. 

Both X-ray absorption and angle-resolved photoelectron spectroscopies, however, are not site-sensitive techniques. In the case of cuprates this means that the role of planar, apical, and \emph{reservoir}-layer oxygens in the charge redistribution cannot be disentangled. The hypothesis is that the different role of these oxygens in determining the properties of cuprates is mapped into a distinct response to charge redistribution. Site specificity can be recovered by looking at core levels via XPS. As shown in Fig.~\ref{fig1}d, suitable overall binding energy resolution allows deconvolving the O\,$1s$ emission into three components with binding energy depending on the bonds and bond coordination. After long debate \cite{Meyer1988, Parmigiani1991, Nagoshi1993, Nagoshi1995, Leiro1995, Qvarford1995, Lele1996}, the components have been attributed to oxygens lying in different planes of the unit cell, as depicted in Fig.~\ref{fig1}d: planar oxygens (Cu-O plane, violet), \emph{reservoir}-layer oxygens (Bi-O plane, grey) and apical oxygens (Sr-O plane, green) \cite{Kuo2018}. Nowadays, thanks to the advent of last-generation free-electron lasers operating at high pulse repetition rate, XPS can be implemented in a pump-probe scheme to access the ultrafast dynamics of core-level states of materials far from the equilibrium condition \cite{Hellmann2012, Kutnyakhov2020, Xian2020}, which we applied in the present study.

\section*{Results and discussion}

\subsection*{O\,$1s$ dynamics}

\begin{figure}[ht]
\centering
\includegraphics[width=0.8\columnwidth]{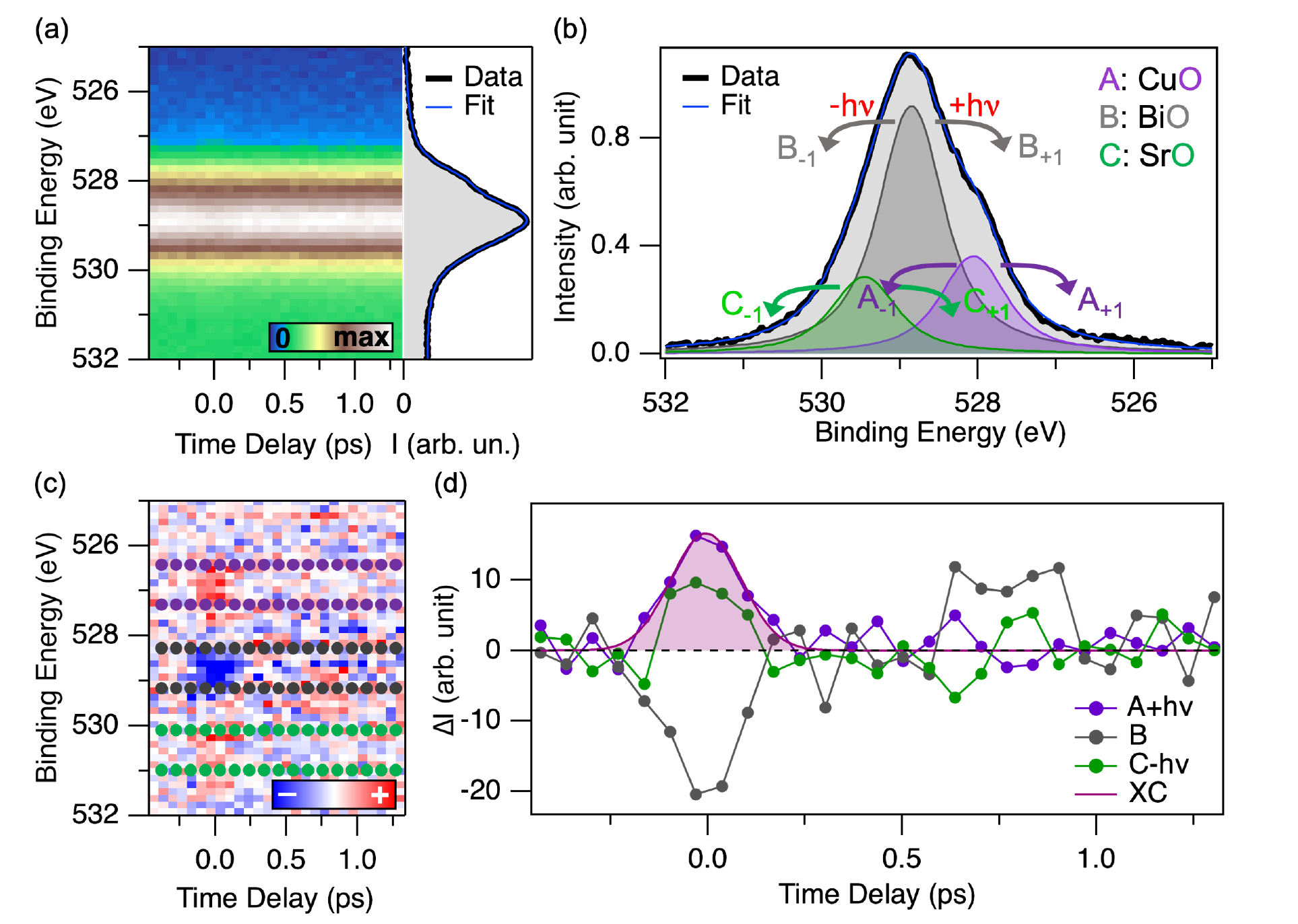}
\caption{(\textbf{a}) Photoemission intensity of the O\,$1s$ emission plotted as a function of time delay and binding energy. The side panel reports the profile of the emission integrated in a 0.2\,ps-wide delay window before time zero. (\textbf{b}) Background-subtracted spectrum integrated in a window of 0.2\,ps around time zero, schematizing the energetics of sideband generation and the nomenclature used in the analysis. (\textbf{c}) Difference between each spectrum of the map in panel \textbf{a} and the average of the spectra in the first 0.2\,ps of the measurement. (\textbf{d}) Traces integrated in the regions denoted by the corresponding colored markers in panel \textbf{c}. The curve superimposed to the violet trace is a Gaussian fit yielding a FHWM of $\sim$0.2\,ps. 
\label{fig2}}
\end{figure}

Figure~\ref{fig2}a shows the intensity distribution of the O\,$1s$ emission as a function of time delay and binding energy. The panel on the right shows the photoelectron distribution in a 0.2\,ps interval before time zero. The spectrum reported is similar to that of the XPS spectra reported in the literature \cite{Parmigiani1991, Meyer1988, Qvarford1996}. The experimental spectrum is fit (blue line) using symmetric functions with different Lorentzian FWHM \cite{Parmigiani1991} and a uniform Gaussian broadening of $\sim 540$\,meV. The parameters obtained for each O\,$1s$ component are reported in Tab.~\ref{tab1}.

\begin{table}[ht]
\centering
\begin{tabular}{|c|c|c|c|c|}
\hline
Component & Plane assignment & Relative Amplitude & Binding Energy (eV) & Lorentzian FWHM (meV) \\
\hline
A       & Planar (Cu-O) &    $0.17 \pm 0.01$    &    $527.97 \pm 0.01$    &    $480 \pm 40$ \\
\hline
B       & Reservoir (Bi-O) &    $1.00 \pm 0.05$    &    $528.84 \pm 0.01$    &    $960 \pm 30$ \\
\hline
C       & Apical (Sr-O) &    $0.14 \pm 0.03$    &    $529.48 \pm 0.04$    &    $660 \pm 90$ \\
\hline
\end{tabular}
\caption{\label{tab1}Summary of the O\,$1s$ parameters extracted from the fit in Fig.~\ref{fig2}a. Amplitudes are reported as fraction of the amplitude of the B component. The errors on the values are derived from the fit uncertainties.}
\end{table}

Before discussing the details of the nonequilibrium evolution of the O\,$1s$ spectra, attention must be payed to a phenomenon known as Laser Assisted Photo-Emission (LAPE). LAPE is well established in gas-phase experiments \cite{Drescher2002, Finetti2017, Maroju2020, Wenthaus2023}. In solids LAPE effects have been predominantly studied on metal surfaces \cite{MiajaAvila2006, Saathoff2008}. Figure~\ref{fig2}b schematizes the processes leading to LAPE in the framework of a phenomenological two-step model. First, the core level electron is emitted by the probe photon into the continuum. Then, the free photoelectron interacts with the infrared field and undergoes absorption or stimulated emission of one or more photons. As a result, smaller replicas of the primary emission appear in the photoemission spectra at binding energy intervals equivalent to steps of $nh \nu$, being $n= \pm 1, \pm 2$ \dots and $h\nu$ the photon energy. The $n$-th order sideband originating from the main emission $X$ ($X = A, B, C$) is thus referenced as $X_{\pm n}$ (Fig.~\ref{fig2}b). The appearance of sidebands in our experiment is clearer in the difference map of Fig.~\ref{fig2}c. The map is obtained by subtracting the spectrum integrated in the first 0.2\,ps of the delay range (side panel of Fig.~\ref{fig2}a) from each of the spectra as a function of the time delay. The blue area around 529\,eV shows that the main emission is losing population because of the rising of the sidebands. Conversely, the region at binding energy distance compatible with one pump photon energy is gaining intensity. In this case, only first-order sidebands are significant. 

The existence of the sidebands is strictly related to the simultaneous presence of photons from the pump and from the probe at the sample position. This can be used as a reference for the origin of the time delay scale, but it is also a reference for establishing the time resolution in our experiment. For this reason, we extract the difference intensity traces from the energy regions marked by round markers in Fig.~\ref{fig2}c, and report them in Fig.~\ref{fig2}d. The photoemission intensity as function of the time delay at the first order sideband is well fit with a Gaussian of $\sim$0.2\,ps FWHM. This gives an estimation of the overall cross-correlation for our experiments, the main contributions coming from the X-ray and laser pulse duration, and the synchronization jitter.

To rationalize the response of the system in terms of lineshape changes, we fit each spectrum of the time series. Each sideband is taken into account as an additional feature with the same lineshape as the corresponding main emission, but a smaller amplitude. Our model thus comprises three oxygen components accompanied by two sidebands. Since each feature is defined by three parameters (amplitude, binding energy, and Lorentzian FWHM), this would yield a total of twenty seven parameters. This number must be reduced also in view of the vicinity of the oxygen components with respect to their width, and of the signal-to-noise ration of the pump-probe measurement. The equal probability of absorption and stimulated emission allows constraining the amplitude of the upper and lower sidebands to be equal. The energy separation of the sidebands is independent of the pump-probe delay, being defined by the pump photon energy. The Gaussian broadening is kept fixed to the equilibrium value \cite{Hellmann2010}. In the time domain, the only constraint applied is on the sideband amplitudes, which are allowed to gain a finite value only for 0.4\,ps around time zero, corresponding approximately to two cross-correlation FWHMs. Finally, Fig.~\ref{fig2}b shows that our model properly fits the O\,$1s$ photoemission data. In particular, it corroborates the assumptions on which the fitting model for the dynamics is built. This yields a total of nine free parameters outside the sideband region (amplitude, binding energy and Lorentzian FWHM) and three additional sideband amplitudes around time zero.

\begin{figure}[ht]
\includegraphics[width=\columnwidth]{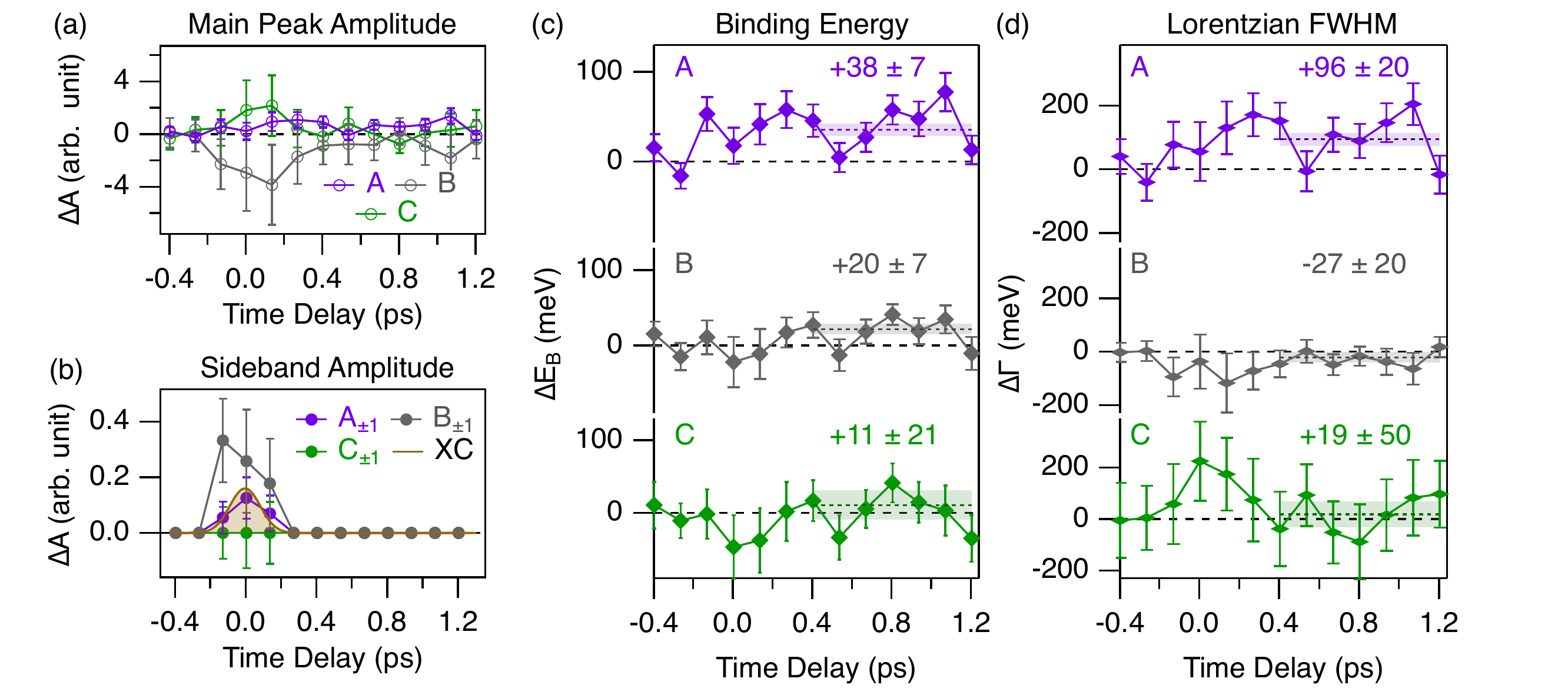}
\caption{Dynamics of the parameters fitting the evolution of the O\,$1s$ components as a function of time delay. The color code indicates the respective component: A (violet, planar), B (grey, \emph{reservoir}), C (green, apical). (\textbf{a}) Amplitude variation for the main emissions. (\textbf{b}) Amplitude variation of the sidebands, equal for upper and lower component and restricted to have finite values only around time zero. The brown curve is the cross-correlation extracted in Fig.~\ref{fig2}d (FWHM $\sim 0.2$\,ps). (\textbf{c}) Binding energy variation of the main emissions. The dashed lines are the average of the points after 0.4\,ps and highlight nonequilibrium values of $+38 \pm 7$\,meV (A), $+20 \pm 7$\,meV (B) and $+11 \pm 21$\,meV (C), the latter being compatible with zero. (\textbf{d}) Lorentzian FWHM variation reported as FWHM. Similarly to panel \textbf{c}, numbers indicate the average value after 0.4\,ps: $+96 \pm 20$\,meV (A), $-27 \pm 20$\,meV (B) and $+19 \pm 50$\,meV (C), the latter being again compatible with zero.
\label{fig3}}
\end{figure}

The evolution of the fitting parameters is shown in Fig.~\ref{fig3} as a variation with respect to the value in the first 0.2\,ps of the delay range. The amplitude of component B (panel a) shows a depletion due to the generation of sidebands. The duration of the negative variation for more than the cross-correlation width indicates that a change in spectral weight is occurring at time zero, and relaxes within half a picosecond. A decrease would also be expected at time zero for the other components, due to sideband generation. At variance, however, the amplitude seems to be increasing, although still being compatible with zero within one standard deviation. A possible explanation is the presence of the sidebands of component B ($\mathrm{B_{\pm 1}}$) which partially overlap with the main emission. This correlation is also demonstrated by the somehow atypical behavior of the sideband amplitude $\mathrm{B_{\pm 1}}$ which, unlike $\mathrm{A_{\pm 1}}$, is only marginally compatible with the expected Gaussian-like behavior (grey and brown curves in Fig.\ref{fig3}b). The sidebands of emission C are hidden in the noise. The maximum amplitude of $\mathrm{A_{\pm1}}$ and $\mathrm{B_{\pm1}}$ is $\sim 0.7 \%$ of the main emission. After 0.4\,ps, all three amplitudes have relaxed back to their equilibrium values. An unambiguous plane dependence is instead displayed by the binding energy and width of the three main features. Component A is the most affected, experiencing a shift to higher binding energies and a broadening already at time zero. Both last for the whole investigated time delay range. The binding energy shift detected is $+38 \pm 7$\,meV whereas the FWHM is estimated to be $+96 \pm 20$\,meV, \emph{i.e.} $\sim22 \%$ broader than the equilibrium value. The modifications of component B are smaller, amounting to $+20 \pm 7$\,meV and $-27 \pm 20$\,meV ($\sim3 \%$ relative to equilibrium). Both the width and the binding energy of emission C have an insignificant average value, suggesting a faster relaxation back to equilibrium. 

\subsection*{Sr\,$3d$ dynamics}

The Sr ions are bound to the apical oxygens, hence lying in the plane adjacent to the Cu-O one. Due to this coordination and geometry, the Sr core levels offer additional information on the role of the apical oxygens in the charge redistribution process.  

\begin{figure}[ht]
\includegraphics[width=\columnwidth]{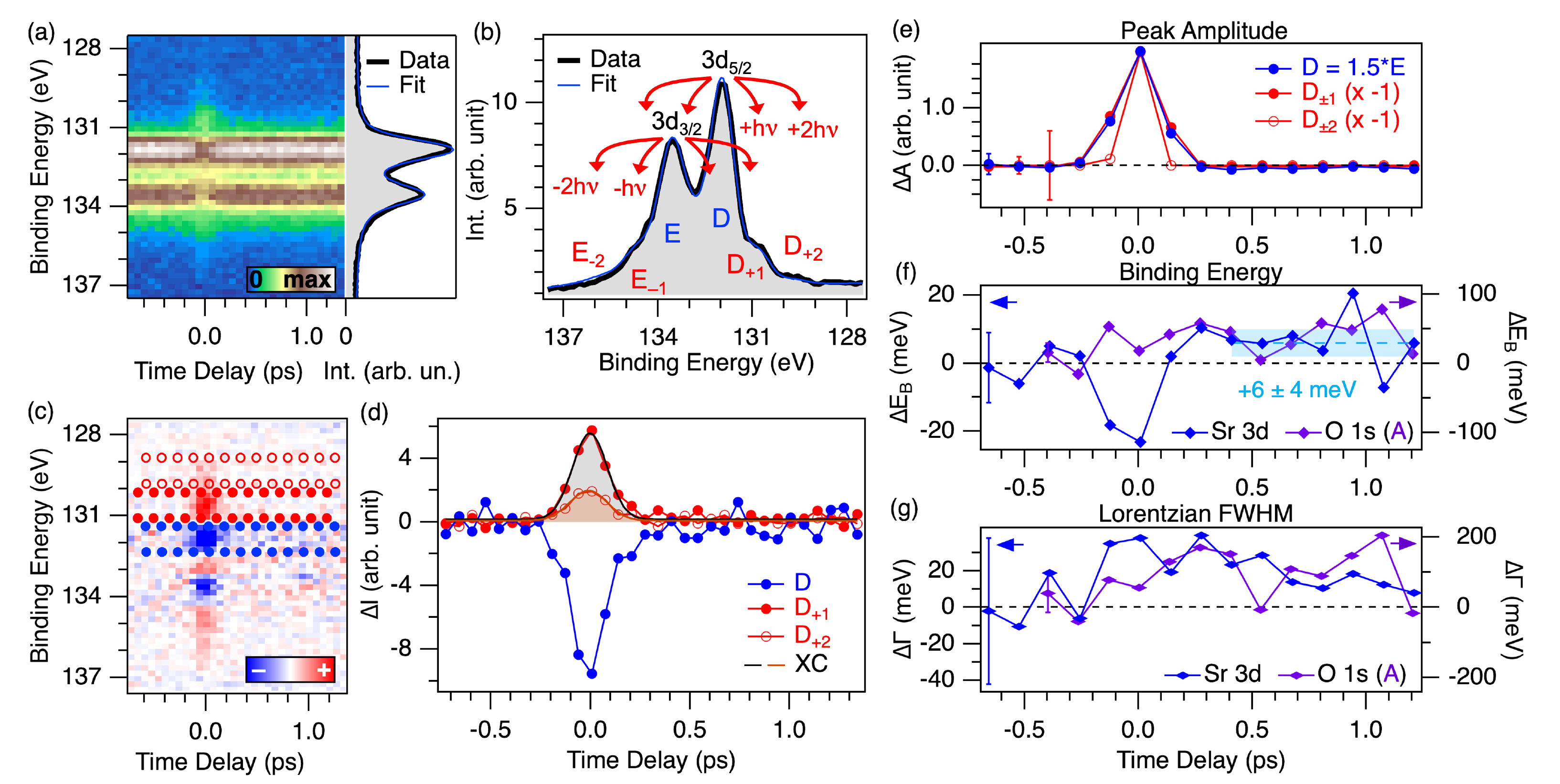}
\caption{(\textbf{a}) Photoemission intensity of the Sr\,$3d$ doublet as a function of time delay and binding energy. Side panel: spectrum integrated in the first 0.4\,ps of the measurement, with the fit conducted as described in the main text (blue line). (\textbf{b}) Spectrum integrated around time zero in a 0.2\,ps window, illustrating the energetics of sideband generation and the nomenclature used in the analysis. In this case, two orders of sidebands are visible. (\textbf{c}) Difference between each spectrum of the map in panel \textbf{a} and the equilibrium spectrum, as a function of the time delay. (\textbf{d}) Traces integrated in the region of the map in panel \textbf{c} denoted by the same marker type. Full blue circles, $3d_{5/2}$ main emission (D). Full red circles, upper first-order sideband (D$_+1$). Empty red circles, upper first-order sideband. Grey and orange lines are the Gaussian fit of the photoemission intensity dynamics around the upper first- and second-order sidebands respectively. All the cases give a FWHM compatible with 0.2\,ps. (\textbf{e})-(\textbf{g}) Dynamics of the fit parameters for the spectra reported in panel \textbf{a} displayed as variation with respect to equilibrium for each time delay. Error bars are similar across the dynamics and are indicated only once in the first points. (\textbf{e}) Amplitude of main feature (filled blue circles), first- and second-order sidebands (filled and empty red circles). (\textbf{f}) Binding energy of the Sr\,$3d_{3/2}$ emission (blue, plotted against the vertical axis on the right) compared with the one of the A component of O\,$1s$ (violet, plotted against the vertical axis on the left). The average after 0.4\,ps indicates a shift of $+6 \pm 4$\,meV of the of Sr\,$3d$ doublet. (\textbf{g}) Lorentzian FWHM of the Sr\,$3d_{3/2}$ emission and of the oxygen A component, plotted as in panel \textbf{f}. 
\label{fig4}}
\end{figure}

Fig.~\ref{fig4}a shows the emission of the Sr\,$3d$ core level as a function of the time delay, with the panel on the right reporting the integration of the emission intensity over the first 0.4\,ps. The equilibrium spectrum resolves well the the $3d_{j = {3/2}}$ and $3d_{j = {5/2}}$ spin-orbit splitting. The Sr\,$3d_{3/2}$ peak (E in Fig.~\ref{fig4}b) is found at $\sim 131.81$\,eV and the doublet splitting amounts to $\sim 1.71$\,eV. The ratio between the area of the two branches is consistent with the theoretical prediction. An overall Gaussian broadening of $\sim 500$\,meV results from the fit, in agreement with the value extracted for the O\,$1s$ components. 

First-order sidebands are clearly visible in the map of Fig.~\ref{fig4}a. After integrating the spectrum around time zero (Fig.~\ref{fig4}b), the second-order sidebands can also be discerned, in contrast to the case of O\,$1s$. The differential map of Fig.~\ref{fig4}c is thus integrated in the energy regions around the main and side features, leading to the evolution displayed in Fig.~\ref{fig4}d. The integration region and the resulting trace are indicated with the same marker. The intensity dynamics in the sideband regions are well fit by a Gaussian with a FWHM of $0.21\pm 0.01$\,ps for the first-order and $0.22\pm 0.02$\,ps for the second-order one. Both are in good agreement with the cross-correlation extracted from the O\,$1s$ spectra. The negative average of the blue signal at later time delays stems from an out-of-equilibrium effect beyond sideband generation. The fitting model needs to comprise one Doniach-Sunjic (see Methods) spin-orbit split feature for the main emission, one for each of the first-order sidebands, and two more for the second-order sidebands. In agreement with what is reported in literature \cite{Hellmann2010}, we assume that the structure of the doublet is not modified by the pump, so that the splitting and the ratio of the spin-orbit split emission is held fixed in the time evolution. Similarly, the asymmetry is fixed to its equilibrium value. In this way, we monitor the time dependence of three parameters: amplitude, binding energy and Lorentzian FWHM. We also impose the amplitude of the upper and the lower sideband to be equal for both orders ($X_{+1,2} = X_{-1,2}$, $X = D, E$) and finite only for two cross-correlations around time zero. The spacing of the sidebands is fixed to one or two pump photons. When fitting the complete dynamics in this framework, the number of free parameters is three outside the time delay region affected by the sidebands (amplitude, binding energy and Lorentzian FWHM of the main doublet) and within it (due to the additional first- and second-order amplitudes). These constraints have been applied to fit the time-zero spectrum in Fig.~\ref{fig4}b, yielding a good agreement. 

Figure~\ref{fig4}e shows the dynamics of main and sideband amplitudes, normalized to the maximum of the main emission to highlight their similarity. The main doublet shows no significant variation except the depletion due to the sidebands. In fact, the dynamics are fully compatible with those of the first-order sideband. The finite amplitude of the second-order sideband has a shorter duration, but we ascribe this to a spurious effect of the fitting procedure due to the low signal-to-noise ratio, since the photoemission intensity variation at the second-order sideband binding energy has the same duration as the first-order one  (full and empty red circles in Fig.~\ref{fig4}d). The amplitude ratio between the first-order sideband and the main emission is of $\sim 18\%$, \emph{i.e.} 30 times larger than the case of oxygen. The binding energy of the peak at time zero displays a shift of $\sim20$\,meV to lower binding energies, presumably of ponderomotive origin (Fig.~\ref{fig4}f) \cite{Jonsson:87}. Comparison with the planar O\,$1s$ component, reported on the same panel against the opposite vertical axis, shows that the two effects have the same sign, but the shift of the Sr\,$3d$ binding energy is one order of magnitude smaller. The width of the feature, although affected by large error bars, also displays a larger change at time zero, which seems to relax in the subsequent dynamics. It is not possible to extract the timescale of this relaxation, since the width has not recovered its equilibrium value within the measured delay range. Based on the available data, a single-exponential decay would give a time-constant of at least 1\,ps. This is qualitatively different from the dynamics of the width of the O\,$1s$ planar component, where no indication of relaxation is found within the measured time delay range. 

\section*{Conclusions}

Our study proves that a charge redistribution in the valence band of Bi2212 induces detectable changes in both O\,$1s$ and Sr\,$3d$ core levels. In the first place, the efficiency of sideband generation is markedly different for the two atomic species. Since second-order sideband generation involves the absorption or emission of two photons, its relative efficiency will scale approximately as the square of the first-order one. For O\,$1s$ the latter is about 30 times smaller than in Sr\,$3d$, so second-order sidebands are expected to be roughly $1000$ times weaker. This rationalizes the absence of second-order sidebands in the O\,$1s$ spectra around time zero. 

Our experiments also prove the capability of detecting the chemical environment-dependent dynamics of oxygen states in Bi2212. In particular, the response of the three components to photoexcitation is quantitatively different. The largest changes are witnessed by the oxygen in the Cu-O plane. Notably, these are the atoms involved in the charge-transfer process described before as the first excitation of the undoped compound. This points to the explanation that the charge redistribution induced by the pump pulse involves predominantly the Cu-O plane, and thus the charge transfer among Cu and O plays a major role. The oxygen binding energy and width, however, do not relax to the equilibrium values within the measured time delay of $\sim 1.4$\,ps. 

The effects experienced by the apical oxygen are not significant in our experiments. This might be due to the larger signals of the neighboring components, which also cause the large uncertainty in the fitting parameters (see error bars in the green curves of Fig.~\ref{fig4}). An indirect proof that the changes experienced by the apical oxygen are small compared to the planar oxygen ones comes from the inspection of the Sr\,$3d$ emission. The lineshape modifications are one order of magnitude smaller than those observed for the planar oxygen. We stress the fact that the sign of the shift is the same for Sr\,$3d$ and O\,$1s$. This allows to rule out pump-induced space-charge as an origin, since an opposite sign would be expected \cite{DellAngela2015, Oloff2014}. Although it is possible that different species in the same plane experience different changes, the evidence points to the fact that the Sr-O plane, and thus also the apical oxygen, is only marginally involved in the charge redistribution induced by the pump. This further confirms that the charge transfer mechanism between Cu and O is the interaction that dominates the relaxation process. 

In conclusion, our experiment has proven the effectiveness of time-resolved XPS in unveiling the element- and coordination-specific dynamics in a cuprate superconductor. The results highlight the peculiar role of the Cu-O plane in the relaxation of the system upon impulsive infrared photoexcitation, and possibly open a new route for studying the dynamics of the many-body interactions in these compounds. 

\section*{Methods}

\subsection*{Sample characterization}

The experiments were performed on an optimally-doped Bi2212 high-quality sample. The superconducting transition temperature, as measured by transport and magnetic (SQUID) methods, was $\sim$ 91\,K \cite{ }. The time-resolved XPS measurements were performed at a base temperature of 100\,K, which is above the superconducting transition temperature, in order to minimize thermal broadening effects in the photoemission spectra.

\subsection*{Experimental setup}

The experiments were performed at the PG2 monochromator beamline at the free-electron laser FLASH at DESY in Hamburg \cite{Martins2006, Gerasimova2011}. To reach the O\,$1s$ core level we used the monochromatized third harmonic at $\sim$ 592\,eV resulting from the fundamental FEL emission at $\sim$ 199\,eV. The presence of the grating induces a temporal stretch of the X-ray pulses up to 150\,fs FWHM. 

The laser source consisted in a Yb:YAG amplifier system delivering pulses at $1030\,$nm with an energy up to $30\,\mu \mathrm{J}$/pulse. The laser source was equipped with a Herriott-type multipass cell for bandwidth broadening, allowing compression of the pulses to $\sim 70$\,fs \cite{Seidel2022}. The temporal jitter between the FEL source and the laser pulses was $\sim 50$\,fs. 

The photoelectrons were detected using a momentum microscope based on a Time-of-Flight design \cite{Schoenhense2015,Kutnyakhov2020}. This allows the simultaneous measurement of the kinetic energy and lateral momentum components. The microscope can also measure the topography of the sample surface with a field of view typically set to $450\,\mu \mathrm{m}$, allowing the measurement of the spot size of both X-ray and laser pulse at the sample position, and monitoring the spatial overlap. The measured spot sizes  were $50\times200\,\mu \mathrm{m}$ for the FEL and $100\times270\,\mu \mathrm{m}$ for the laser: the smaller X-ray spot size guarantees an homogeneous optical pumping of the system in the probed area. The laser intensity was set to $\sim 2.4\,\mu \mathrm{J}$/pulse, which implies a fluence of $\sim 1\,\mathrm{mJ/cm^{2}}$, suitable to mitigate pump-induced vacuum space-charge effects \cite{Schonhense2021}. 

The pump-probe configuration was obtained by coupling an optical laser pulse into the propagation path of the  X-ray pulse in a quasi-collinear configuration. The pump and probe pulses impinged on the sample at an angle $\theta = 68^{\circ}$ with $p$-polarization, and the optical pulse was stabilized to maintain the spatial overlap with the X-ray pulses.

\subsection*{Analysis of the spectra}

The calibration of the energy and time delay axes was conducted as described in previous works \cite{Dendzik2020, Curcio2021, Xian2020}. The photoelectron spectra were fit with a sum of Doniach-Sunjic functions \cite{Doniach1970} above a Shirley background \cite{Shirley1972}. The Doniach-Sunjic functions were convoluted with a Gaussian to account for several broadening sources, \emph{i.e.} photon pulse and monochromator bandwidth, thermal broadening, and possible residual space charge. The distribution of photoelectrons in each energy bin was assumed to be Poissonian, and hence the uncertainty on the number of counts in a bin estimated as its square root. This was taken into account as a statistical error in the fit and hence contributes to the uncertainty on the extracted parameters. 

To fit the pump-probe measurement, the fitting procedure was repeated for each spectrum at a variable time delay. The initial parameters were obtained from the fit of the spectrum integrated before time zero; the results of the fit of the $n$-th spectrum were then used as initial parameters to fit the $(n+1)$-th spectrum in the time series. 


\section*{Acknowledgements}

We acknowledge support by the scientific and technical staff of FLASH as well as Holger Meyer and Sven Gieschen from the University of Hamburg for support of the HEXTOF instrument.  D.K., N.W. and M.H. acknowledge the funding by the SFB 925 ``Light Induced Dynamics and Control of Correlated Quantum Systems'' - 170620586 (project B2). 

The work at BNL was supported by the US Department of Energy, oﬃce of Basic Energy Sciences, contract no. DOE-sc0012704.

\section*{Author contributions statement}

F. Pressacco and F. Parmigiani conceived the experiment; G.G. and R.J.C. grew and characterized the samples; D.P., D.K., L.W., M.S., N.W., M.H., G.B., W.B., F.C., F. Parmigiani and F. Pressacco conducted the experiments: D.P. and F. Pressacco analyzed the results; D.P., F. Pressacco and F. Parmigiani drafted the manuscript with major input from F.C. and W.B.; all the authors extensively discussed the results and reviewed the manuscript. 




\section*{Additional information}

The authors declare no competing interest. 

\end{document}